\documentclass[letterpaper, 10 pt, conference]{ieeeconf}  

\IEEEoverridecommandlockouts                              

\usepackage[USenglish]{babel}

\usepackage{todonotes}
\usepackage{multirow}
\usepackage{tabularx}
\usepackage{diagbox}
\usepackage{rotating}
\usepackage{makecell}
\usepackage{subcaption}
\usepackage{enumerate}

\usepackage{color, colortbl}
\definecolor{Gray}{gray}{0.9}

\usepackage{fancyhdr}

\fancypagestyle{specialfooter}{%
  \fancyhf{}
  
  \fancyfoot[L]{\footnotesize This work was accepted by the International Conference on Systems, Man, and Cybernetics, Kuching, Malaysia, 2024.\linebreak
  © 2024 IEEE. Personal use of this material is permitted.  Permission from IEEE must be obtained for all other uses, in any current or future media, including reprinting/republishing this material for advertising or promotional purposes, creating new collective works, for resale or redistribution to servers or lists, or reuse of any copyrighted component of this work in other works.}
}

\newtheorem{definition}{\textbf{Definition}}

\title{\LARGE \bf
Perspectives-Observer-Transparency -- A Novel Paradigm for Modelling the Human in Human-To-Anything Interaction Based on a Structured Review of the Human Digital Twin
}

\author{Nils Mandischer$^{1,\dag}$, Alexander Atanasyan$^{2,\dag}$, Michael Schluse$^{2}$, Jürgen Roßmann$^{2}$, and Lars Mikelsons$^{1}$
\thanks{* This work was funded by the StMWK in the ``Augsburg AI Production Network'' as part of the High-Tech Agenda Plus.}
\thanks{\dag Both authors contributed equally to this research.}
\thanks{$^{1}$ N. Mandischer and L. Mikelsons are with the Chair of Mechatronics at University of Augsburg, 86159 Augsburg, Germany.
        {\tt\small nils.mandischer@uni-a.de}}%
\thanks{$^{2}$ A. Atanasyan, M. Schluse, and J. Roßmann are with the Institute for Man and Machine Interaction at RWTH Aachen University, 52074 Aachen, Germany.
        {\tt\small atanasyan@mmi.rwth-aachen.de}}%
}

\begin{document}

\maketitle
\thispagestyle{empty}
\pagestyle{empty}

\begin{abstract}
Modern modelling approaches fail when it comes to understanding rather than pure supervision of human behavior. As humans become more and more integrated into human-to-anything interactions, the understanding of the human as a whole becomes critical. In this paper, we conduct a structured review of the human digital twin to indicate where modern paradigms fail to model the human agent. Particularly, the mechanistic viewpoint limits the usability of human and general digital twins. Instead, we propose a novel way of thinking about models, states, and their relations: Perspectives-Observer-Transparency. The modelling paradigm indicates how transparency -- or whiteness -- relates to the abilities of an observer, which again allows to model the penetration depth of a system model into the human psyche. The split in between the human's outer and inner states is described with a perspectives model, featuring the \emph{introperspective} and the \emph{exteroperspective}. We explore this novel paradigm by employing two recent scenarios from ongoing research and give examples to emphasize specific characteristics of the modelling paradigm.
\end{abstract}

\setcounter{footnote}{2}
\thispagestyle{specialfooter}

\section{Introduction}
Imagine a smart city of the future. You get off a plane and are guided to your local destination. You are tired of sitting from the long journey. That's why instead of sending an autonomous car, the city's intelligent dispatcher guides you through personalized information along your pathway. You meet other pedestrians, which are on their own journey, interacting with the smart city. At the destination, the door opens autonomously and a lift is already waiting to take you on. An intelligent assistant welcomes you and a robot assists you in the kitchen. All this is not a vague vision, but already a central cornerstone of diverse innovation strategies~\cite{Burdett.2015,acatech.2011}. However, this not only involves the city on the high-level but also cars driving in the city, traffic lights controlling pedestrians and vehicles, or billboards showing personalized advertisement, among many other cyber-physical devices the human may interact with. And this does not even stop at cyber-physical devices, but continues with vegetation, static building assets, or other humans, which may function as a partner in an human-to-anything (H2X) interaction. A central aspect of those innovation strategies is, that the smart and digitized city and all its accommodated entities are enabled to understand the human's needs and desires. This requires a digital representation of the human agent and the surrounding entities. While real entities are commonly digitized as digital twins which establish a digital representation of the real world, such representations are not readily available for the human. Human Digital Twins (HDTs) are already subject to research but lack to represent the human as a whole in a systematic manner. Instead they use rather mechanistic viewpoints on the real human, trying to simulate behavior without an elaborate understanding of the inner states. Therefore, the human's inner state is still rather opaque in modern human-system integration (HSI) making it hard to develop a detailed understanding of the human agent.

In this paper, we analyze the state of the art in human modelling, namely HDTs (Section~\ref{sec:review}). Within, we show that modern HDTs are rather mechanistic in nature and that novel modelling paradigms are required to close the gap from pure supervision to understanding of the human agent. We emphasize this gap by discussing two examples where mechanistic model views fail (Section~\ref{sec:scenarios}). Based on this, we introduce a novel way of modelling the interdependence of inner and outer states of the human agent: \textbf{Perspectives}, depicting inner and outer states, which are subject to an \textbf{Observer} with specific abilities to bring \textbf{Transparency} to these, particularly, inner states (Section~\ref{sec:perspectives}). These aspects are addressed in the Perspectives-Observer-Transparency modelling paradigm, which we discuss along exemplary H2X systems (Section~\ref{sec:applications}).

Summarizing, the main contributions of this paper are:
\begin{itemize}
    \item Structured review of the Human Digital Twin
    \item Introduction of the novel Perspectives-Observer-Trans\-parency modelling paradigm for systems involving human agents
    \item Exemplary uses of the modelling paradigm based on recently implemented HSIs
\end{itemize}

\section{Structured Review of the Human Digital Twin}
\label{sec:review}
To find similarities and indicate potential gaps in human modelling, we conduct a structured review of the HDT. We analyze the HDT in context of mechanistic systems and its ability to model and simulate an anthropomorphic system through understanding of the relation of outer (observable/physical) and inner (non-observable/non-physical) system states. To give a better base for discussion, we first introduce a definition of mechanistic systems (\mbox{Section~\ref{ssec:review_mechanistic}}) and our chosen method for the literature review (\mbox{Section~\ref{ssec:review_method}}). Afterwards, we discuss the literature on the HDT and indicate challenges to overcome (Section~\ref{ssec:review_hdt}).

\subsection{Mechanistic Systems}
\label{ssec:review_mechanistic}

A structured mechanistic system is defined as a system which is an aggregation of individual parts that contribute to the functionality of the whole with their individual purpose or task~\cite{Moser.2021}. Given this definition, the human is also an agglomeration of individual parts, e.g., limbs or organs. The mechanistic view observes these parts by means of their geometry, motion, and general physical or physiological behavior. The human, however, is more than the mere observation of the physical behavior of their parts. A mechanistic system is inanimate, it has no inner state (in the sense of this work). While in mechanistic system models, the interpretation of the outer state typically delivers the required insight of their inner workings, it is the true inner states that define the human -- an animate system. For the following discussions, we use the terms model and system according to VDI~3633~\cite{VereinDeu.2018} (Definitions~\ref{def:system} and \ref{def:model}).

\begin{definition}
    \emph{A \textbf{system} is a set of interrelated elements that are separated from their environment.}
    \label{def:system}
\end{definition}

\begin{definition}
    \emph{A \textbf{model} is a simplified reproduction of a planned or existing system with its processes in a different conceptual or physical system.}
    \label{def:model}
\end{definition}

\subsection{Method}
\label{ssec:review_method}
\begin{figure}[tb]
    \centering
    \includegraphics[width=0.35\textwidth]{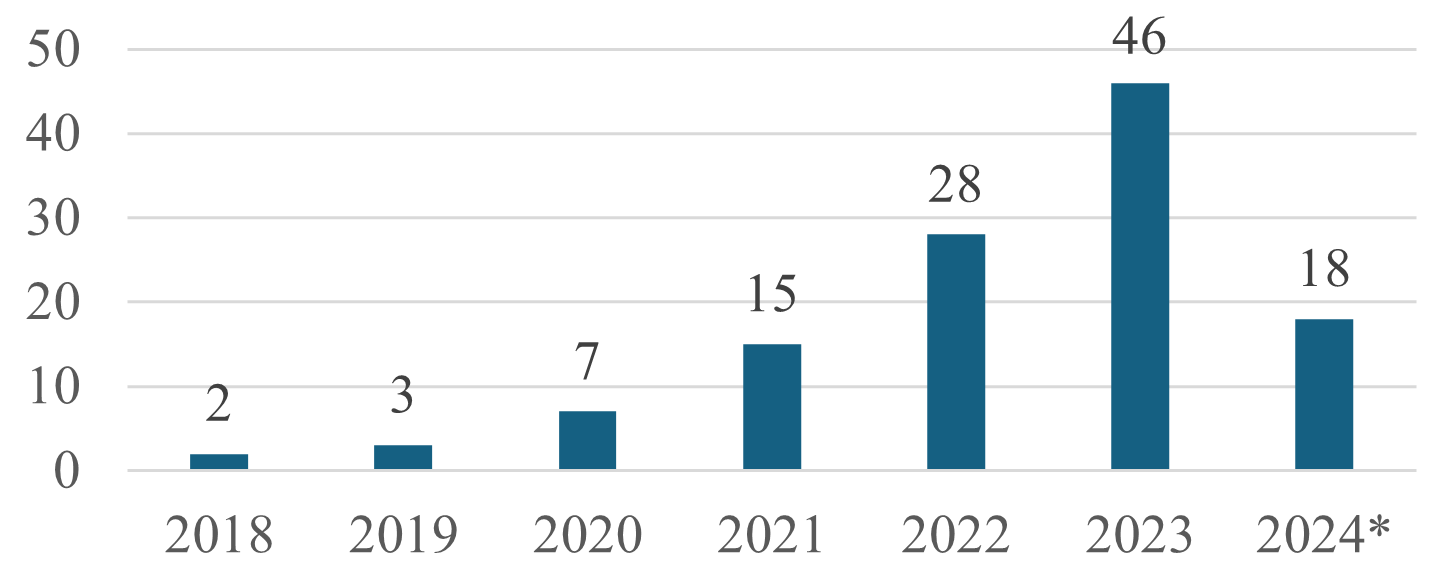}
     \caption{Yearly number of publications included as per described criteria between 2018 and March 2024 (*).}
    \label{fig:num_publications}
\end{figure}

To explore the broad scope of the multifaceted perspective on modelling the human, our structured review examines literature across technical and medical databases: The \textit{ACM Digital Library}, \textit{IEEE Xplore}, \textit{Science Direct}, \textit{PubMed}, and the four \textit{APA PsycNet} databases. Our searches (9~Boolean search strings\footnote{\scriptsize1: "Digital Twin" AND ("human aspects" OR "psychological aspects"); 2: "Digital Twin" AND ("inward perspective" OR "emotions" OR "motivations"); 3: "Digital Twin" AND ("outward perspective" OR "social interaction" OR "environmental interaction"); 4: ("Human Digital Twin" OR "Artificial Agent") AND ("internal goals" OR "personal goals"); 5: "Digital Twin" AND ("interconnection" OR "integration") AND ("inward perspective" OR "outward perspective"); 6: "Human Digital Twin" AND ("cognitive models" OR "emotional models"); 7: "Digital Twin" AND ("agent-based modeling" OR "behavioral modeling"); 8: "Digital Twin" AND "multi-perspective" AND ("human factors" OR "artificial intelligence"); 9: "Human Digital Twin" AND ("Architecture" OR "Methodology")}) aim to cover defined aspects of inner and outer perspectives of human and artificial agents, along with these perspectives' interconnections. Notably, search term~5 yields no results across all databases. The database search took place in March 2024 and had no filtering by date. Initially, 2893 results were imported into \emph{Rayyan} for deduplication, screening, and filtering, leaving 2140 results after deduplication and the removal of full proceedings. Filtering for ``human digital twin'' (broad matching) narrowed the field to 166 publications. Through screening, these were reduced to 119 due to the exclusion of papers lacking a focus on human or artificial agent modelling. This detailed screening targeted the identification of recent approaches in structuring the human digital twin for simulation purposes and H2X interactions. We did not engage in additional snowballing. As the low amount of just two results in 2018 and rapid super-linear increase in the following years suggests (see Figure~\ref{fig:num_publications}), the HDT can be considered a nascent but rapidly developing approach quickly gaining interest. We made the review database publicly available at~\cite{database}. 

\begin{figure*}[t!]
    \centering
     \begin{subfigure}[t]{0.49\textwidth}
         \centering
         \includegraphics[width=.87\textwidth]{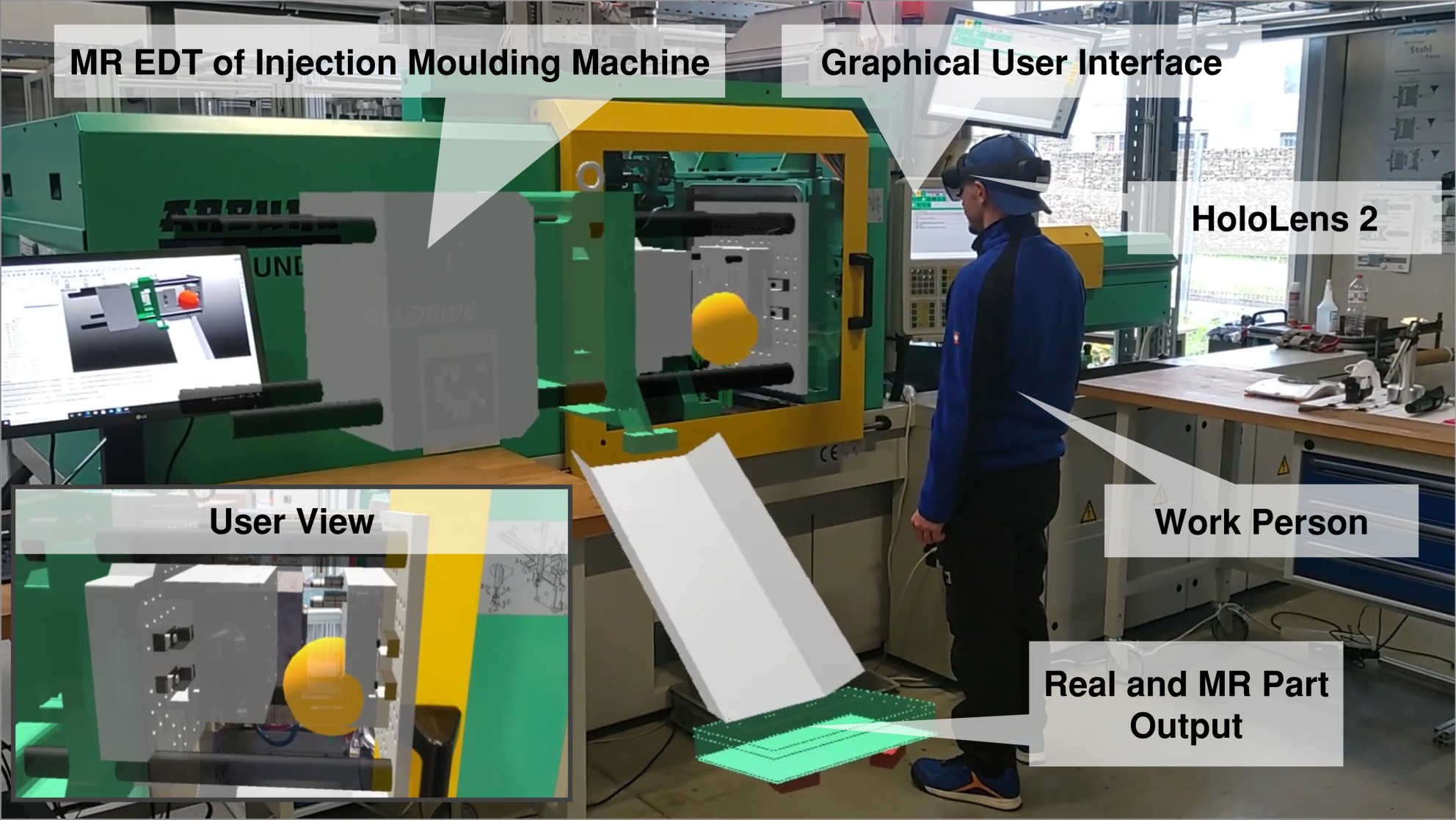}
         \caption{Human-supervisor interaction: The supervisor shows the work person a virtual version of the injection molding machine.}
         \label{sfig:scenario_1}
     \end{subfigure}
     \hfill
     \begin{subfigure}[t]{0.49\textwidth}
         \centering
         \includegraphics[width=.87\textwidth]{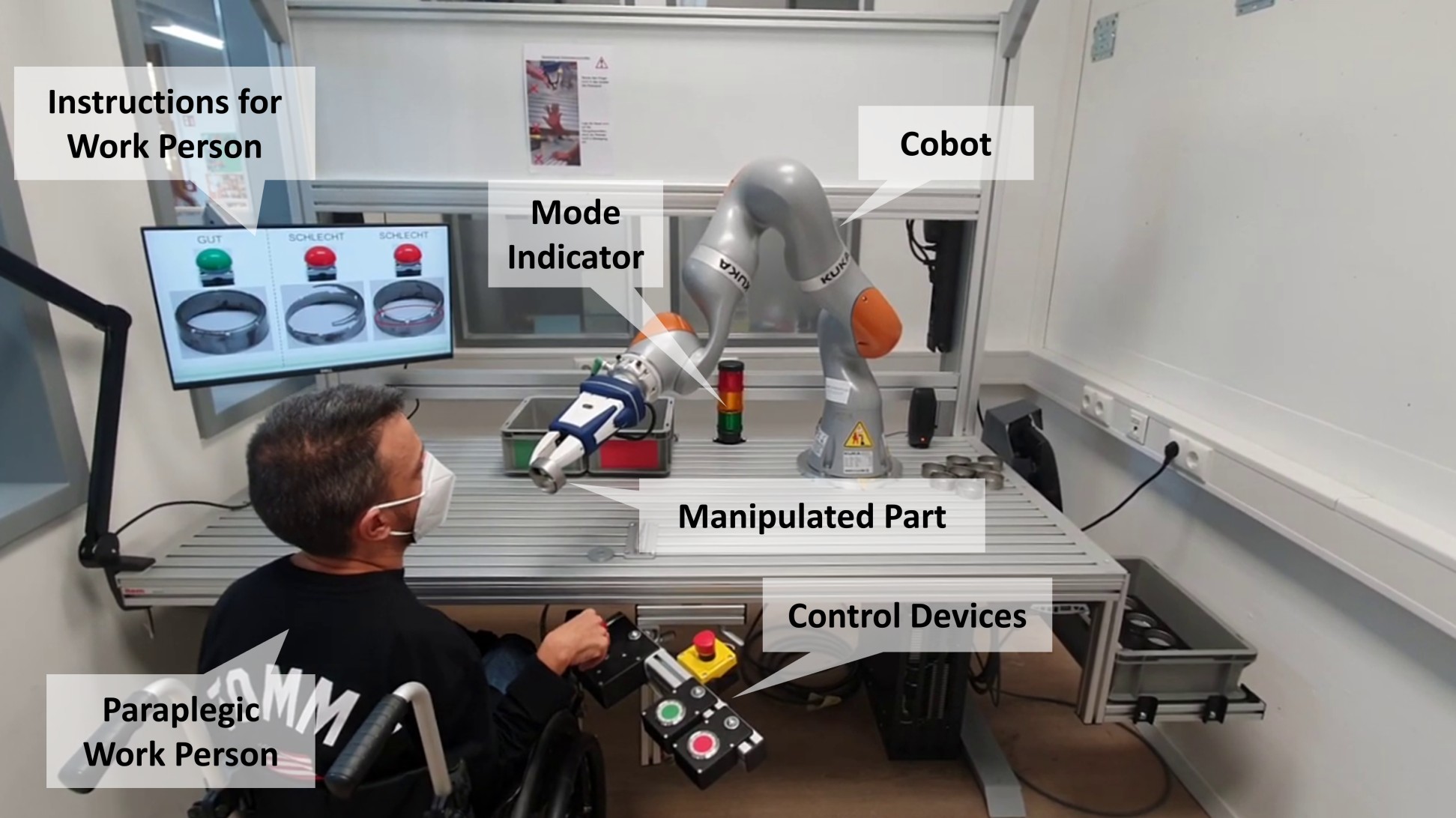}
         \caption{Human-robot interaction: The human manually controls the robot to rotate a part (\cite{NextGeneration}, annotated).}
         \label{sfig:scenario_2}
     \end{subfigure}
    \caption{Two exemplary scenarios in manufacturing featuring a human-integrated system. It becomes obvious in both scenarios that human and machine are not fully integrated but merely co-exist.}
    \label{fig:scenarios}
\end{figure*}

\subsection{Human Digital Twins}
\label{ssec:review_hdt}


In developing a methodological approach to human-machine interaction, the HDT emerges as a key conceptual framework. Stemming from the work on Digital Twins (DTs) by Grieves~\cite{grieves_2014}, the HDT seeks to replicate the multifaceted nature of human beings in the digital realm. Some core aspects motivating an extension of the DT approach beyond technical assets and towards the human are: (1) their role as the linchpin for looking at real assets' historic states, properties, and behavior, (2) the representation of their current state, and (3) views onto possible futures of the real twin, when extended by appropriate simulation approaches. While for the technical DT a broader common understanding emerges, the HDT is a more recent concept, apparently lacking unified understanding, architectures, or implementations.

The main fields of application for HDTs are diverse with examples found in industrial~\cite{davila-gonzalez_human_2024, mordaschew_human_2024}, medical~\cite{okegbile_human_2023, subasi_digital_2024, chen_networking_2024, li_dtbvis_2023}, and urban planning~\cite{liu_towards_2023, gnecco_digital_2023}, reflecting the concept’s adaptability to various domains. In the industrial domain, HDTs facilitate process simulation and workplace optimization, evolving from the focus of Industry~4.0 to the more human-centric alignment of Industry~5.0. This shift emphasizes a closer, more natural interaction between humans and machines, prioritizing human well-being through task matching and ergonomic considerations. The medical field leverages HDTs for personalized healthcare, enabling better simulation of therapeutic effects and the development of organ-specific DTs. The HDT concept is also applied to enhance human well-being, leading to digital interfaces for preliminary design processes and improving interaction with physical spaces. The HDT concept aligns with the Operator~4.0 framework~\cite{CPG23, Romero.2016}, with an outlook towards Operator~5.0. Various architectures have been proposed for \mbox{HDTs~\cite{davila-gonzalez_human_2024, okegbile_human_2023, naudet_preliminary_2023, wang_human_2024, soldatos_7_2021,locklin_architecture_2021}}, often focusing on specific domains or presenting broad, high-level perspectives. These architectures consider the distinction between physical, emotional, cognitive, or mental aspects of the human experience. A significant number of models aim to infer non-observable states, such as emotional and mental states, comfort, or general well-being -- often calibrated by correlating collected data with self-reported values. However, a systematic distinction and connection between observable outer and non-observable inner states is notably lacking, indicating a need for a more structured approach to inference.

Several contributions stand out: Davila-Gonzalez et al.~\cite{davila-gonzalez_human_2024} propose a high-level architecture focusing on mental, emotional, and physical domains, yet lack explicit connections between them. Subasi and Subasi~\cite{subasi_digital_2024} aim to extend beyond mechanistic models, emphasizing ethics and individualization within therapeutic optimization. However, they lack a concrete implementation strategy. Sharma and Gupta~\cite{sharma_leveraging_2024} argue for the superiority of ``cognitive'' DTs, that can intelligently adapt and perform tasks, over passive models. Mordaschew et al.~\cite{mordaschew_human_2024} consider both, physical and mental, aspects of the human as structural blocks without modeling their interrelation. Chen et al.~\cite{chen_networking_2024} acknowledge the complex interrelations within HDTs and introduce an architecture that emphasizes networking and the integration of AI methods with a narrow exemple application. Finally, Wang et al.~\cite{wang_human_2024} take the broadest stance of all reviewed architectures. They propose a generic architecture addressing human needs within human-machine interaction. They outline their expression in various roles and point out layers pertinent to Industry~5.0 and the existence of various models relevant to the HDT like physical, cognitive, psychological, collaborative, and organisational. However, they do not address detailed relationships between the data and various models in the HDT, and lack a hierarchy of levels of understanding.

\section{Scenarios with Limitations of Mechanistic Digital Twins}
\label{sec:scenarios}
To enrich the gap identified in Section~\ref{sec:review} with examples, we discuss two scenarios, that show typical limitations in modelling the human in H2X scenarios. Even though recent HDT architectures already model outer and inner states of the human and their interconnection, they are still mechanistic by nature, which is reflected in the examples.

\subsection{Human-Supervisor Interaction in Industrial Learning}

\subsubsection{Scenario Description}
This scenario focuses on the idea of allowing human learners to make errors when working in industrial manufacturing (Figure~\ref{sfig:scenario_1}). Wherever possible, a superordinate supervision layer prevents consequences of erroneous actions in the real world, but shows them in a Mixed Reality (MR) environment. In the exemplary use case of injection molding machine operation, when a dangerous state is reached (e.g., the process settings would cause high wear to the machine), the supervision layer, based on defined rules, stops the machine's process before damage occurs and starts the machine simulation in MR. This approach makes it possible to leverage errors to better understand the nature of work processes and all involved objects, where human teachers would otherwise interfere, e.g., to avoid injury or damage~\cite{Atanasyan.2020}.

\subsubsection{Design of Digital Twin}
The concept has been realised using Experimentable Digital Twins (EDTs)~\cite{Schluse.2018}. EDTs track the present state of the current training scenario and use their simulation-based capabilities to express the behavior of, e.g., industrial machines, hence, predicting and assessing the consequences of human actions and allowing to experience the future consequences of errors.

\subsubsection{Limitations}
Aside from EDTs that are able to represent present and potential future states of the technical systems, models are missing which explain how the learner perceives the process and the work environment, and how they derive knowledge from this perception. This also requires an understanding of the process which the learner attempts to acquire. This model, which is a typical human inner state, enables the supervisor to detect and track actions and, based on this knowledge, detect the erroneous actions and help the learner to improve their work. Only detecting a dangerous machine state and stopping the machine is not sufficient in this scenario.

\subsection{Human-Robot Teaming for Inclusion of People with Disabilities into Manufacturing Processes}

\subsubsection{Scenario Description}
In this scenario, people with disabilities (PwD) are assisted in manual work tasks by collaborative robots (Figure~\ref{sfig:scenario_2}). The paradigm used is cooperation, i.e. robot and human share a workspace and task; central aspect is the assignment of tasks to the agents.

\subsubsection{Design of Digital Twin}
For task allocation, on the one hand, the capabilities of the human are determined using an occupational analysis tool (see Weidemann et al.~\cite{Weidemann.2022}). The analysis is based on ergonomics and occupational medicine. On the other hand, the process is modelled to deduce requirements. Both, capabilities and requirements, are then matched and the tasks which the human is unable to perform are re-allocated to the robot. The robot is programmed accordingly.

\subsubsection{Limitation}
There is no shared model of the goal and how to reach this goal. This leads to a mutual split in individual tasks. Both agents fulfil complete process steps only; there is no physical interaction. Through the pre-defined task allocation, adapting to in-situ needs of the PwD is not possible, e.g., if the PwD gets tired or inattentive --- information that is typically provided by inner states. Therefore, the system fails in more complex and new situations. While human-centric, the matching tool uses a mechanistic view on the human-robot system by only evaluating observable states, e.g., the execution of a task or test. There is only minimal assessment of the inner states given in the mental capabilities, which are evaluated only at singular occurrences by professionals. Consequently, this procedure is not properly suited for longer work processes.

\section{Perspectives-Observer-Transparency Model}
\label{sec:perspectives}
As the two exemplary scenarios underline, the used digital twins of the humans are too restrictive in their abilities to properly integrate the human into the technical system. The approaches only model the human (and the technical system)  ``as seen from the outside''. When humans participate in a technical system, this perspective is not sufficient. In fact, Vinciarelli et al.~\cite{Vinciarell.2015} identified i.a. that approaches for analysis and synthesis of human behavior lack a deeper understanding of the human (``how people [...] give meaning to their experiences'') and that the interaction of technical systems with humans remains shallow. Consequently, mechanistic system design will inevitably create a barrier which cannot be overcome within this paradigm.

\begin{figure}[tb]
    \centering
    \includegraphics[width=\linewidth]{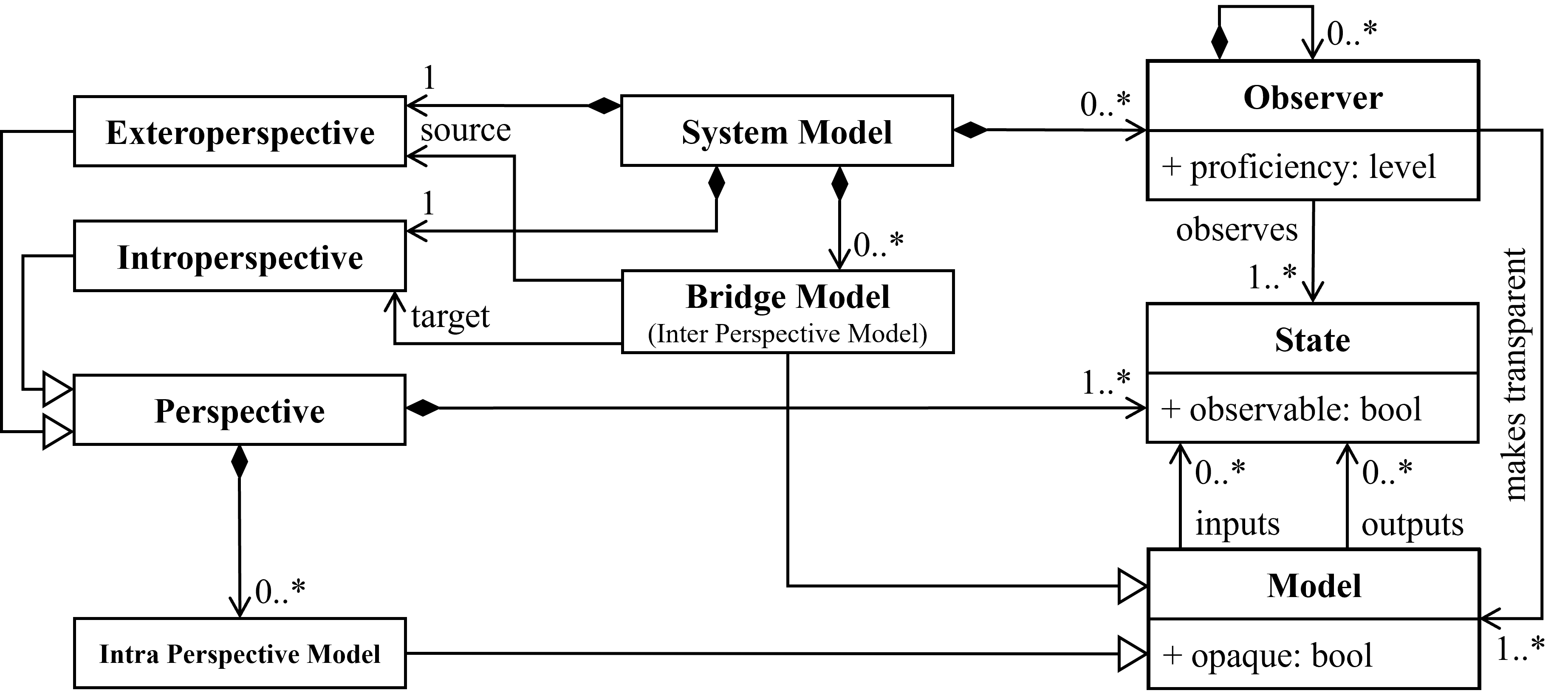}
    \caption{UML diagram of the classes in the Perspectives-Observer-Transparency modeling paradigm.}
    \label{fig:poto_uml}
\end{figure}

In the following, we introduce a novel way to think about modelling digital twins of humans in the context of HSI: Perspectives, Observers, and Transparency. Perspectives are the viewpoints an agent may take on the human agent (Section~\ref{ssec:perspectives}). We assume that the entire system model (including sub-models and states) is initially opaque~(\mbox{Section~\ref{ssec:transparency}}). Observers (\mbox{Section~\ref{ssec:observer}}), then, use their knowledge to bring transparency to this opaque system model. Each observer uses technical measures (Section~\ref{ssec:level}) required to assess states in the system model representing the human agent, e.g., fatigue, stress, or engagement. The Perspectives-Observer-Transparency modelling paradigm (Figure~\ref{fig:poto_uml}) will help to overcome the barriers of mechanistic digital twins and streamline the understanding of the human.


\subsection{Perspectives}
\label{ssec:perspectives}
The major difference of the common technical system and the human agent lies within the human's inner states, that are mostly not understandable (not measurable or traceable) by modern modelling techniques and algorithms. We emphasize this by dividing models of a system into two perspectives -- the \emph{Introperspective} and the \emph{Exteroperspective} -- according to Definitions~\ref{def:introperspective} and \ref{def:exteroperspective}, respectively.
\begin{definition}
    \label{def:introperspective}
    \emph{A state is part of the \textbf{introperspective} if it represents a \underline{non-physical} state. A model is part of the introperspective if its inputs and outputs are only connected to states of the introperspective.}
\end{definition}

\begin{definition}
    \label{def:exteroperspective}
    \emph{A state is part of the \textbf{exteroperspective} if it represents a \underline{physical} state. A model is part of the exteroperspective if its inputs and outputs are only connected to states of the exteroperspective.}
\end{definition}

\begin{figure*}[t!]
    \centering
    \includegraphics[width=.82\textwidth]{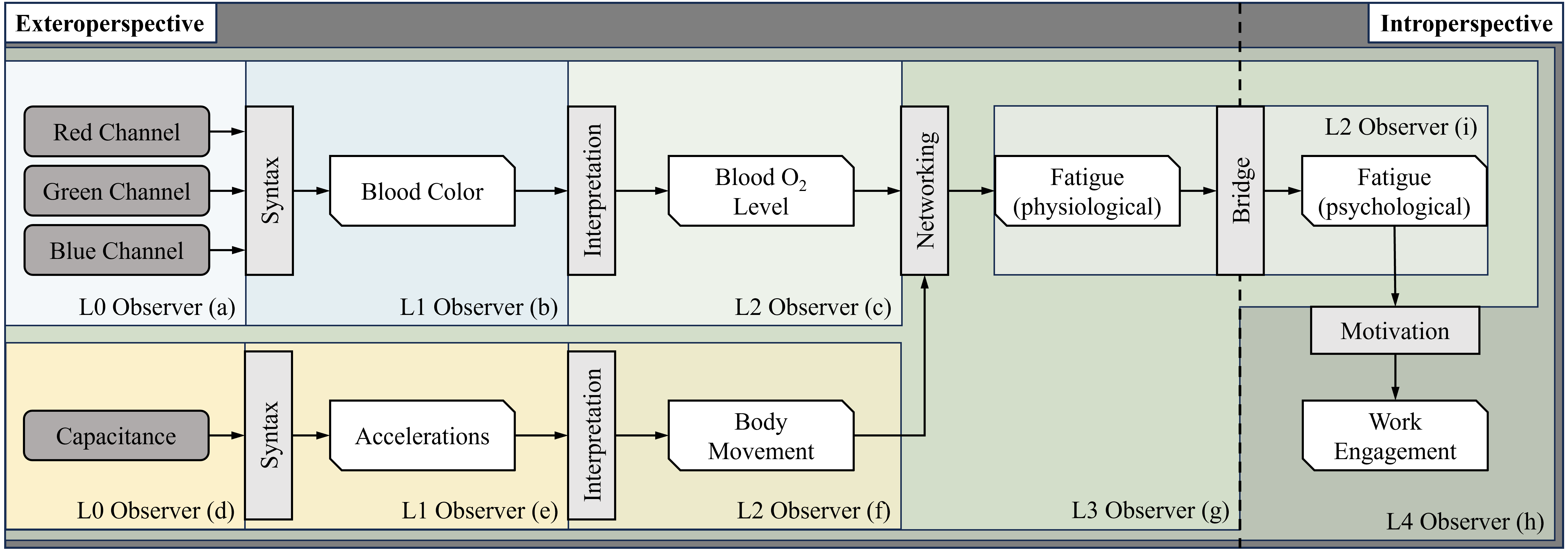}
    \caption{Perspectives-Observer-Transparency model with an exemplary system model based on Austin et al.~\cite{Austin.2020} and Secher, Seifert, and van Lieshout~\cite{Secher.2008} to analyze work engagement based on diverse intermediate states. Rounded boxes are direct measures, cut boxes are states, and normal boxes are models; arrows indicate flow of information; L: Level. The oxygen-level is only the interpretation of the blood color which is derived from the color channels of an RGB image ($a$, $b$, $c$). The same is true in an IMU sensor used for detecting body movement ($d$, $e$, $f$). Both information are used to detect physiological fatigue in the \emph{exteroperspective}, which functions as an indicator for psychological fatigue in the \emph{introperspective}. The bridge model between them is essentially an L2 observer ($i$) embedded in L3 observer $g$. The \emph{introperspective} psychological fatigue is then used as an indicator in a motivation model to evaluate work engagement as part of L4 observer $h$.}
    \label{fig:landscape}
\end{figure*}

The two perspectives define a mutual split between the models and states. Figure~\ref{fig:landscape} depicts an exemplary distribution of models and states within the \emph{exteroperspective} and the \emph{introperspective}. What becomes obvious is that some states (or at least similarly named states) exist in both perspectives. This is, as some aspects of the human agent are both, physiological (physical) and psychological/mental (non-physical). Per Definitions~\ref{def:introperspective} and \ref{def:exteroperspective}, a model cannot exist in both perspectives. For this purpose, we define a bridge model, which is essentially an interpretation of the physical towards the non-physical aspect of the corresponding state. Can, Mahesh, and Andr\'{e}~\cite{Can.2023} emphasize this relation of physiological and psychological factors in human behavior analysis. Note that the bridge model itself with its connected states is essentially part of a level~2 observer (cf. \mbox{Section~\ref{ssec:level}}).

\begin{definition}
    \label{def:bridge}
    \emph{A model is a \textbf{bridge model} if its inputs and outputs are connected to at least one state in each perspective.}
\end{definition}

\subsection{Opacity vs. Transparency}
\label{ssec:transparency}
Methods to model certain aspects of a technical system are commonly categorized as black or white~\cite{LoyolaGon.2019}. The terminology is inherently binary, hence, we will use the terminology ``opacity/opaque'' and ``transparency/transparent'', instead. Both terminologies are interchangeable. 

Ashby~\cite{Ashby.1957} terms opacity as the relation of an object and its observer. In fact, there is no true transparency in any real object but it is merely the knowledge and experience of the observer which makes the object transparent. Glanville~\cite{Glanville.1982} underlines this aspect of opacity by indicating that even though a system is considered transparent, it is actually opaque on the outside. Both theories are effectively true if the system can be continuously detailed out. E.g., if we pull an object with a rope, it is common understanding that the rope transmits the force, while the true nature of the system lies within the interaction of subatomic entities. As a model is made for a purpose, including framework boundaries and a set level of detail~\cite[p.~31]{Cobelli.2019}, this overarching understanding of opacity is not feasible. For this paper we consider a model to exist only within certain system boundaries and with a certain level of detail. Thus, we use Definitions~\ref{def:opaque} and~\ref{def:transparent} for opacity/transparency. For transparency, we partially adapted the definition by Cellier and Kofman~\cite{Cellier.2006}, who define opacity as the inability to induce correcting actions based on the interpretation of the system (see aspect (b) in Definition~\ref{def:transparent}).
\begin{definition}
\label{def:opaque}
    \emph{A model is considered \textbf{opaque} if it cannot be understood \underline{but} by the pure interpretation of its outputs.}
\end{definition}

\begin{definition}
\label{def:transparent}
    \emph{A model is considered \textbf{transparent} if (a) it describes all relevant aspects of the modelled system and its interactions within priorly set system boundaries with permissible error \underline{and} (b) its inner workings are comprehensible for an observer with sufficient system knowledge.}
\end{definition}

\begin{table*}[t!]
    \centering
    \caption{Formalization of the \emph{Knowledge Stairway} with abilities of an observer required to reach proficiency in specific knowledge. Higher levels include prior capability requirements. The level of uncertainty rises with the observer proficiency.}
    \begin{tabular}{p{.015\textwidth}|p{.09\textwidth}|p{.45\textwidth}|p{.13\textwidth}|p{.18\textwidth}}
        ~ & \textbf{Proficiency} & \textbf{Required Abilities} & \textbf{Target Perspective} & \textbf{Examples}\\
        \hline
        \textbf{0} & Symbols & perceive/measure symbols or physical entities & exteroperspective & photocell, button\\
        \textbf{1} & Data & establish syntax between symbols; typically not effected by uncertainty & exteroperspective & RGB~camera,~IMU\\
        \textbf{2} & Information & interpret data including time series of data; typically impacted by uncertainty & exteroperspective & Luenberger observer, pulse oximeter\\
        \textbf{3} & Knowledge & interconnect information from different sources and establish a knowledge network & border & fatigue models, diagnostics\\
        \hline
        \rowcolor{Gray}
        \multicolumn{5}{c}{\textbf{Embed into an experimentable framework}}\\
        \hline
        \textbf{4} & Understan\-ding actions & establish goal models that motivate observed states & introperspective & motivation models, Pavlovian conditioning \\
        \textbf{5} & Competence & find the correct goal which motivates the observed states & introperspective & Human as observer
    \end{tabular}
    \label{tab:observer}
\end{table*}

\subsection{Observer}
\label{ssec:observer}
While we use different definitions for transparency and opacity, the definition of Ashby~\cite{Ashby.1957} ultimately leads to the challenges within the mechanistic digital twins: the technical system has no sufficient knowledge and experience to turn the opacity of the human agent into transparency. By this, the human remains opaque even though there exist models which \textbf{interpret} their behavior. To solve this challenge, we introduce an observer according to Definition~\ref{def:observer}.
\begin{definition}
\label{def:observer}
    \emph{An \textbf{observer} is a technical or human system with specific abilities and knowledge of another human, technical system, or a combination of both, whose perspective can be used for a system analysis.}
\end{definition}
The observer takes the role of an entity to which we can bind the capability to convert opacity into transparency as part of a model's attributes (cf. Figure~\ref{fig:poto_uml}). According to Definition~\ref{def:observer}, the technical systems in Section~\ref{sec:scenarios} are insufficient observers. If no observer is deployed, the system model stays opaque. Within the opaque system, models are ordered into the perspectives. Applying an observer generates a transparent region within the opaque system representation (see Figure~\ref{fig:landscape}), i.e. opaque models and states become transparent. By this, the transparent region represents the part of the system which is comprehensible by the observer, i.e. which they can \textbf{understand}. Everything outside the transparent region is subject to interpretation or unknown. Therefore, the observer is a representation of a technical solution for implementing transparent models. A common modelling technique to implement such a solution is by applying (model-agnostic) interpretability techniques (e.g., \mbox{Riberio, Singh, and Guestrin~\cite{Ribeiro.2016}}), which aim at training transparent models with the output data of opaque models, hence, offering interpretation of the opaque model by analyzing the inner workings of the transparent model. Within the transparent regions, an observer may use inference techniques to connect different states throughout multiple models, e.g., in form of a POMDP~\cite{Spaan.2012} or a Bayesian network~\cite{Jordan.2008}.

\begin{figure}[t]
    \centering
    \includegraphics[width=.42\textwidth]{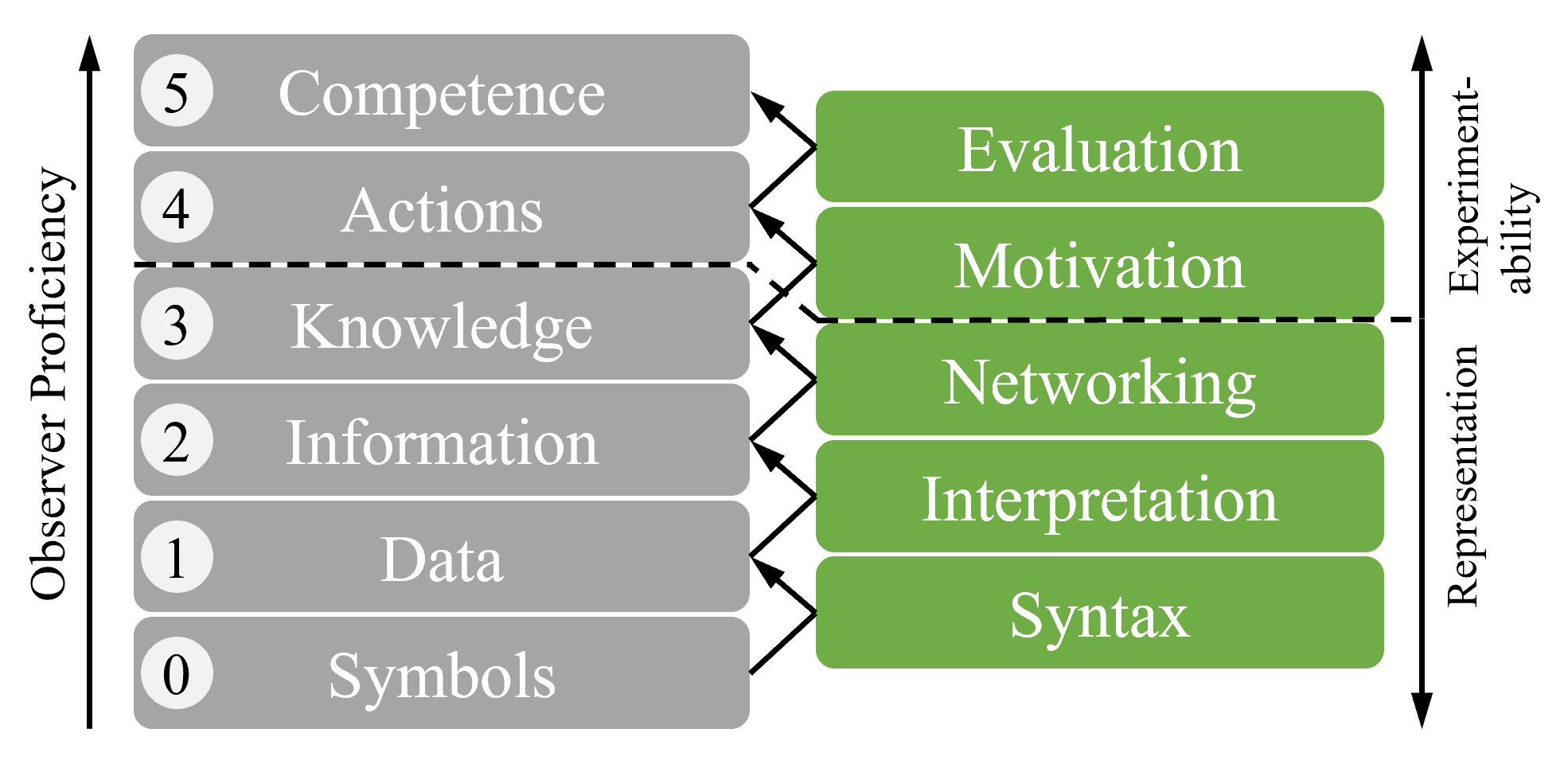}
    \caption{Adaptation of the \emph{Knowledge Stairway} for the Perspectives-Observer-Transparency modelling paradigm. The dashed line represents where experimentability is required for higher proficiency, which coincides with the border between \emph{introperspective} and \emph{exteroperspective}.}
    \label{fig:stairway}
\end{figure}
\subsection{Observer Levels}
\label{ssec:level}
To indicate the ability of an observer to understand the system, we employ the \emph{Knowledge Stairway} (from German ``Wissenstreppe'') by North~\cite{North.2016}. The \emph{Knowledge Stairway} indicates operative actions to leverage the knowledge proficiency towards competitiveness of a company. However, most aspects transport well to knowledge generation on the human agent. We adapt the \emph{Knowledge Stairway} in Figure~\ref{fig:stairway} and annotate proficiency levels (L), which are further detailed in Table~\ref{tab:observer}. In L0 to L3, the observer creates knowledge networks to bring transparency to the border region of the \emph{introperspective} (cf. Figure~\ref{fig:landscape}). This is the very limit a pure knowledge representation may achieve. To achieve higher levels of proficiency towards competence in human understanding, the observer needs to benefit from H2X interaction in form of experimentability, e.g., as experimentable digital twin~\cite{Schluse.2018}. Through interaction with the human agent, the system representation is given an exploration space to uncover human states. Exploration allows for selecting between multivariate state values which are initially equally probable. In context of digital representations, experimentability enables exploration through (real-time) simulation of probable historic, current, and future states. This is a key aspect if human actions and goals shall be assessed. We removed the competitiveness stage in North's model, as we do not aim for marketable value.
When transporting the observer idea to human-human interaction, it becomes obvious that the human is an exceptional observer of other human agents. There are multiple works indicating that human teams perform best when they communicate implicitly (e.g., Shah and Breazeal~\cite{Shah.2010}), which requires a well-defined understanding of the \emph{introperspective}, particularly, motivation and intention of behavior. An example may be found in the emergency room, where the surgical nurse presents the tools required for the proceeding step without the surgeon asking for them. This leads to a natural and exceptionally efficient working symbiosis (e.g., Mulligan~\cite{Mulligan.2016}).

\section{Applications of the Perspectives-Observer-Transparency Model}
\label{sec:applications}
In this section, we apply and discuss the Perspectives-Observer-Transparency modelling paradigm in relation to the mechanistic systems described in Section~\ref{sec:scenarios}.

\begin{figure*}[t!]
    \centering
     \begin{subfigure}[t]{\textwidth}
         \centering
         \includegraphics[width=.82\textwidth]{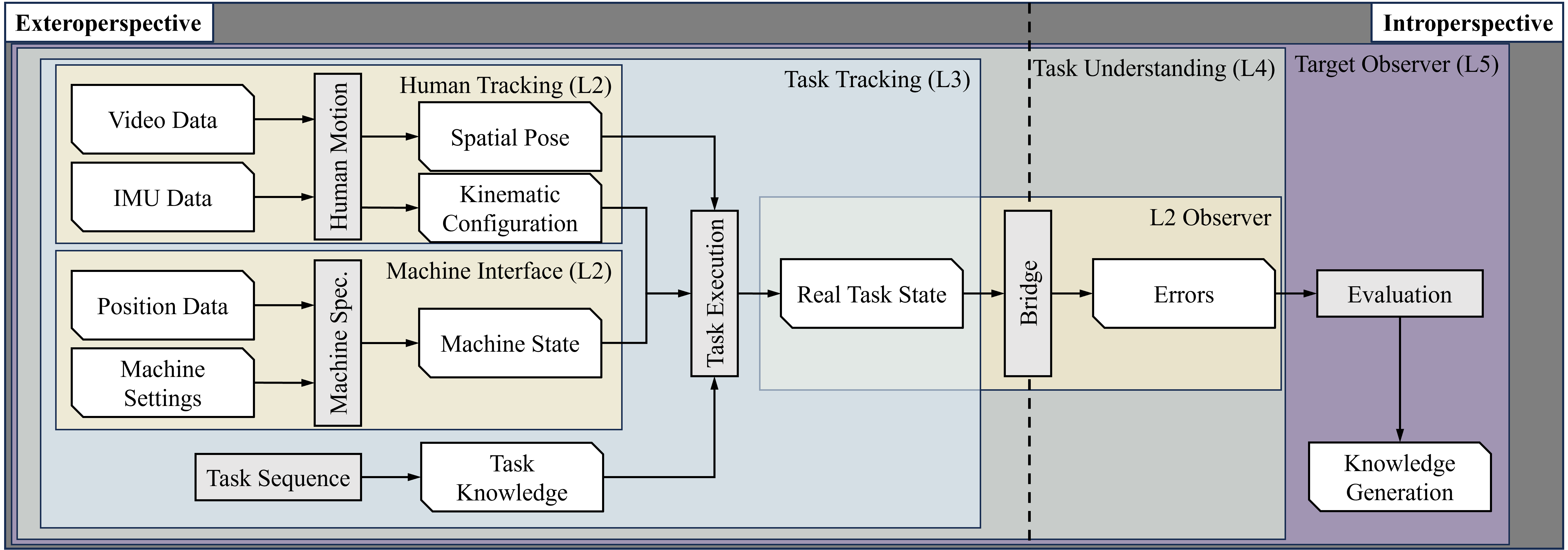}
         \caption{The analysis of semantically labelled errors that were made during task execution allows for a more directed and individualized learner debriefing. Errors are modelled as the difference between nominal and actual task execution.}
         \label{sfig:validation_fedinar}
     \end{subfigure}
     \hfill
     \begin{subfigure}[t]{\textwidth}
         \centering
         \includegraphics[width=.82\textwidth]{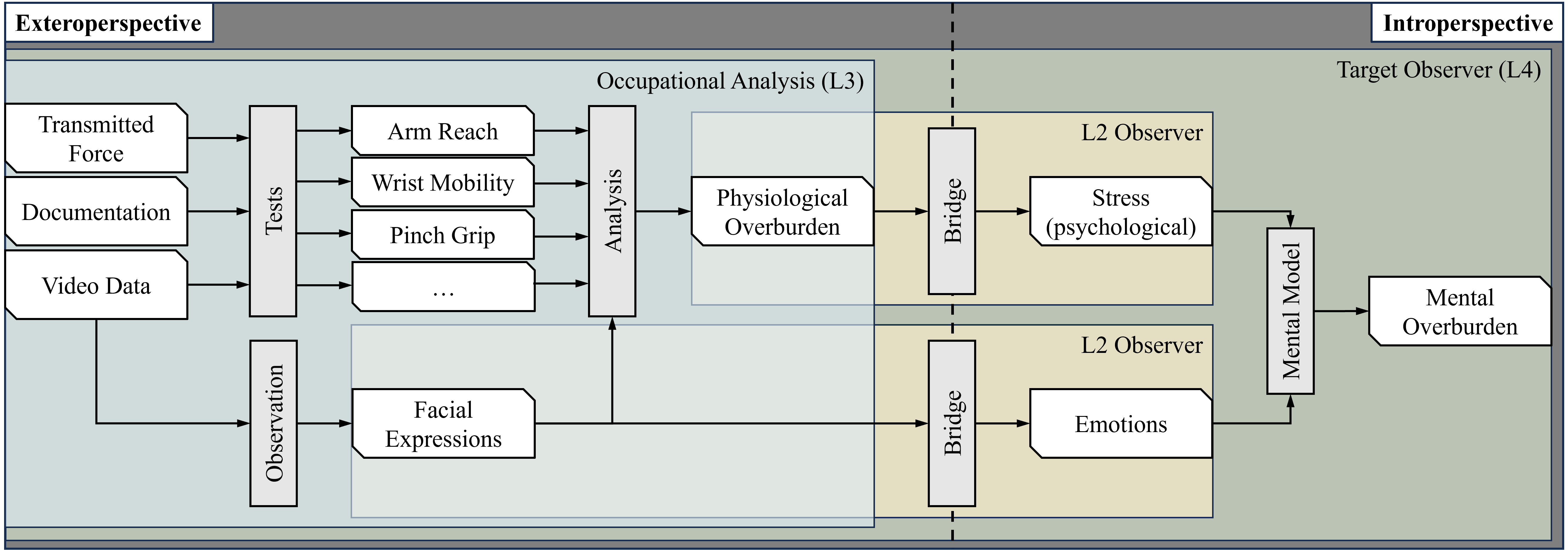}
         \caption{The occupational analysis by the physician is improved by incorporating \emph{introperspective} models for detection of mental overburden.}
         \label{sfig:validation_nextgen}
     \end{subfigure}
    \caption{Perspectives-Observer-Transparency model for the improved twins of Section~\ref{sec:scenarios} / Figure~\ref{fig:scenarios}.}
    \label{fig:validation}
\end{figure*}

\subsection{Recipe}
Applying the Perspectives-Observer-Transparency modelling paradigm is rather simple, even though the implications are complex. Given a system model with set system boundaries and its underlying models and states, first:
\begin{enumerate}
    \item Order models and states into \emph{introperspective} and \emph{exteroperspective}.
    \item Assume all perspectives' space to be opaque.
    \item Set transparency regions for given observers, e.g., a H2X system already implemented or to be evaluated.
\end{enumerate}
If after these steps, a mismatch between the observers and the desired level of transparency is determined, perform following steps:
\begin{enumerate}
    \setcounter{enumi}{3}
    \item Find observers with sufficient transparency.
    \item Implement higher-level observers by employing the operative action according to the \emph{Knowledge Stairway}, potentially including new models and states.
\end{enumerate}
The major challenge in deploying this recipe lies within the implementation of the observers, which becomes gradually more complex, the higher the level. Note that not all potential observers are readily available, which is majorly influenced by the readiness of specific models, e.g., there is no generalist model for human desire. Further note that if singular models are missing, the other knowledge relations may be modelled as part of other (lower level) observers (cf. Figure~\ref{sfig:validation_nextgen}).

\subsection{Improvements to the Human-Supervisor Interaction}
\label{ssec:improvement_fedinar}

\subsubsection{Applying the Paradigm}
The current approach is machine-centred in regard to the work process. The human's motivation is not directly part of the analysis, but it is approximated by their motion and a spatial semantic relation of body and machine parts. Hence, all models are part of the \emph{exteroperspective} and the supervisor layer is a L3 observer.

\subsubsection{New Findings}
To overcome the limitation and analyze how the human agent learns the work process, the observer needs not only to employ motivation models but also evaluation models to understand the human's way of learning. Therefore, a L5 observer would be required to tackle the scenario profoundly. The topic of human learning in itself is manifold~\cite{Jarvis.2012} and selecting the correct learning models is complex. Based on the EDT (as experimentable framework), we implemented a task sequence model and a cognitive model, displaying the learner's perception of the work process (see Figure~\ref{sfig:validation_fedinar}). The cognitive model functions as a motivation model, allowing assessment of knowledge generation in the human agent. The task knowledge is based on the task sequence model. These models elevated the supervision layer only to a L4 observer, but we already observed improvements over the pure mechanistic system: the learners' individual task knowledge can be understood semantically and respective feedback was generated in a partially automated way. However, with a L5 observer even better results are expected, particularly in the seamless transition from real to virtual interaction between the human agent and the machine.

\subsection{Improvements to the Human-Robot Team}
\label{ssec:improvement_nextgen}

\subsubsection{Applying the Paradigm}
The majority of analysis tasks are currently performed by a human, e.g., an occupational physician. While this observer is theoretically able to operate on L5, they limit themselves to testing procedures and video analysis, which is closer to a L3 observer in this case. This is in line with the occupational analysis tool, that performs interpretation of networked knowledge. The technical system operates on an even lower level (L2), as it only interprets direct inputs, e.g., through joysticks or buttons. Therefore, no observer is able to react to the human's true in-situ needs and the robot is unable to react to emotions or engagement.

\subsubsection{New Findings}
A first step towards better fashioned-out collaborative task taking is to elevate the level of the technical observer, i.e. aligning the knowledge of the human and artificial agent. By this, the technical observer is able to take similar conclusions as an occupational physician (Figure~\ref{sfig:validation_nextgen}). The main benefit of the technical over the human observer is the direct interface to the robot controls, hence, better opportunities for the robot to act. A second step would be to elevate the maximal observer level in all observers to leverage better knowledge of the \emph{introperspective} to better react to the human's engagement in the work process, particularly, mental overload in case of people with disabilities. Such elevation of the observer level would require mental models like emotion recognition (e.g., Can, Mahesh, and Andr\'{e}~\cite{Can.2023}). The experimentable framework is already in place in form of a human-robot interaction, but a solution was not implemented for this scenario due to the sheer complexity of the required models. Instead this is subject to ongoing research (see Mandischer et al.~\cite{Mandischer.2023}).

\section{Conclusion}
In this work, we first performed a structured review of the human digital twin, concluding that the majority of system models only cover the human ``as seen from the outside''. We emphasized this fact by indicating challenges in two scenarios employing mechanistic digital twins. To overcome these challenges, we introduced the novel Perspectives-Observer-Transparency modelling paradigm, which defines perspectives (\emph{introperspective}/\emph{exteroperspective}) for models and their connected states, that are opaque in themselves, as well as observers, that generate transparent regions within the perspectives. All aspects were flanked with comprehensive examples and definitions. We, finally, showed how the modelling paradigm may be applied in the two exemplary scenarios and indicated derived actions, which were already partially implemented and evaluated. To leverage the full potential of the modelling paradigm, more scenarios need to be looked at to explore, particularly, edge cases and to refine the definitions given in this paper. 

\bibliographystyle{IEEEtran}
\bibliography{references_citavi, references_review_v0}

\end{document}